\documentclass[12pt]{iopart}
\usepackage[dvips]{graphicx}
\begin{document}
\jl{9}

\title[Continuous variable teleportation and nonclassicality]%
      {Quantum channel of continuous variable teleportation
       and nonclassicality of quantum states}

\author{Masahiro Takeoka*\dag{}, Masashi Ban\ddag{} and Masahide Sasaki*\dag{} \footnote[3]{E-mail:psasaki@crl.go.jp}}
\address{* Communications Research Laboratory,
         4-2-1 Nukui-Kitamachi, Koganei, Tokyo 184-8795, Japan \\
	\dag{} CREST, Japan Science and Technology \\
	\ddag{} Advanced Research Laboratory, Hitachi Ltd.,
		 1-280 Higashi-Koigakubo, Kokubunji, Tokyo 185-8601, Japan\\}

\begin{abstract}
Noisy teleportation of nonclassical quantum states via a two-mode squeezed-vacuum state is studied with the completely positive map and the Glauber-Sudarshan $P$-function.
Using the nonclassical depth as a measure of transmission performance, we compare the teleportation scheme with the direct transmission through a noisy channel.
The noise model is based on the coupling to the vacuum field.
It is shown that the teleportation channel has better transmission performance than the direct transmission channel in a certain region.
The bounds for such region and for obtaining the nonvanished nonclassicality of the teleported quantum states are also discussed.
Our model shows a reasonable agreement with the observed teleportation fidelity in the experiment by Furusawa {\it et al.} [Science {\bf 282}, 706 (1998)].
We finally mention the required conditions for transmitting nonclassical features in real experiments.
\vspace{3mm}

\end{abstract}

\section{Introduction}

Applications of the entanglement of quantum states, {\it e.g.} quantum teleportation or quantum dense coding, have attracted a lot of interests in the field of quantum information technology because of their potential transporting capabilities of quantum or classical informations.

Quantum teleportation was originally proposed for teleporting a qubit (a spin-$\frac{1}{2}$ particle or a two-level system) state by using the two-dimensional EPR state \cite{Bennet1993} and then generalized into the teleportation of continuous variables by using continuously entangled states \cite{Vaidman1994}.
More practical scheme of the continuous variable teleportation has been proposed \cite{Braunstein1998,milburn1999,enk1999} in which a two-mode squeezed-vacuum state is employed as an entangled state.
The first experiment was performed by using quantum optical fields \cite{furusawa1998}.
A two-mode squeezed-vacuum was generated from a sub-threshold optical parametric oscillator (OPO) and the teleportation of a coherent state was demonstrated with the fidelity of $0.58 \pm 0.02$ which was higher than the predicted classical limit of $1/2$.

Although the teleportation of a coherent state has several advantages in practical point of view, quantum teleportation of continuous variables is fundamentally able to transfer arbitrary unknown quantum states including a variety of nonclassical states.
The nonclassical states, such as squeezed states, Fock states and superpositions of the coherent states (Schr\"{o}dinger-cat states), play an important role in quantum information theory and thus it seems important to study about the teleportaion of nonclassical properties of the quantum states.
The nonclassicality of the quantum state is straightforwardly revealed in its $P$-function representation.
The nonclassical state is defined by the state whose $P$-function takes negative values or is singular more than delta-function such as a derivative of delta-function \cite{walls1994}.
Nevertheless, theoretical works studied so far have almost been based on the other representations, such as the Wigner function representation \cite{Braunstein1998}, and mostly concentrated on the analysis of the fidelity of teleportation \cite{Braunstein1998,loock2000,hofmann2000} which can not provide the precise information about the transferred properties of the nonclassicality of quantum states.
Furthermore, to discuss the applicability of the teleportation to quantum communications, the whole process of the teleportation should eventually be treated as a quantum channel from quantum information theoretical point of view.

Lee {\it et al.,} \cite{Lee2000} discussed teleportation of the nonclassical properties of the continuous variable quantum states based on noisy quantum channel models.
They pointed out that any nonclassical properties can not be transferred without nonseparability of the squeezed-vacuum pair and also showed that the teleportation was more fragile than the thermally noised direct transmission channel.

Meanwhile the parameter called ``nonclassical depth'' has been proposed to estimate the strength of the nonclassicality \cite{Lee1991,Lee1992,Lutkenhaus1995}. 
In this paper, with the help of the nonclassical depth, we investigate that how much nonclassicality can be transferred by the noisy teleportation of continuous variables and if the capability of the teleportation is better than that of the direct transmission or not.
In our model, it is assumed that the noise comes from the coupling between the system and an environment in the vacuum state, which is commonly encountered in optical quantum communication networks.
We reveal that the transfer capability of nonclassicality by the teleportation strictly depends on the degree of the two-mode squeezing, the loss of the channel, and the strength of the initial nonclassicality of the quantum state to be teleported.
It is shown that the teleportation channel has better transmission performance than the direct transmission in a certain region.

In section 2, the general procedure of the continuous variable quantum teleportation in lossy channels is briefly summarized, and then formulated by the completely positive (CP) map in section 3.
The quantum state, fidelity and the nonclassical depth for the teleportation channel are derived by using the $P$-function representation in section 4 and 5.
These are then compared to those of the direct channel model in section 6.
Section 7 is the conclusion of this paper.

\section{Protocol of the teleportation of continuous variables}

Quantum teleportation of continuous variables \cite{Braunstein1998,milburn1999,enk1999,furusawa1998} between the sender, Alice, and the receiver, Bob, is performed by sharing a two-mode squeezed-vacuum state $|\Psi^{AB}_{\rm SV} \rangle$ given by
\begin{eqnarray}
\label{eq-2-1}
	\vert\Psi_{\mathrm{SV}}^{AB}\rangle
	&=\exp\left[r(\hat{a}^{\dagger}b^{\dagger}
	-\hat{a}\hat{b})\right]\vert 0^{A}\rangle
	\otimes\vert 0^{B}\rangle \nonumber\\
	&=\sqrt{1-\lambda^{2}}\sum_{n=0}^{\infty}\lambda^{n}
	\vert n^{A}\rangle\otimes\vert n^{B}\rangle,
\end{eqnarray}
where $\hat{a}$ $(\hat{b})$ and $\hat{a}^{\dagger}$ $(\hat{b}^{\dagger})$ are the bosonic annihilation and creation operators for the mode A (B), respectively.
$|n^{A} \rangle$ and $|n^{B} \rangle$ are the photon-number eigenstates of the mode A and B, respectively, and the parameter $\lambda$ is defined by $\lambda=\tanh r$.
For the sake of simplicity, the squeezing parameter $r$ in \Eref{eq-2-1} has been assumed to be positive through this paper.
The mode A and B are assigned to the modes for Alice and Bob, respectively.

In a realistic situation, since the environment inevitably influences the two-mode squeezed-vacuum shared by Alice and Bob, the pure squeezed-vacuum state is turned into the mixed state and the quantum entanglement is degraded.
A state change of the quantum states induced by the environment is fully described by a completely positive (CP) map \cite{davies1976,kraus1983,ozawa1984}.
Thus the mixed quantum state $\hat{\rho}^{AB}_{\rm SV}$ shared by Alice and Bob is represented by the following expression:
\begin{equation}
\label{eq-2-2}
	\hat{\rho}_{\mathrm{SV}}^{AB}
	=\left(\hat{\mathcal{L}}^{A}\otimes
	\hat{\mathcal{L}}^{B}\right)
	\vert\Psi_{\mathrm{SV}}^{AB}\rangle
	\langle\Psi_{\mathrm{SV}}^{AB}\vert,
\end{equation}
where $\hat{\cal L}^A$ and $\hat{\cal L}^B$ are the CP maps for the mode A and B, respectively, and we consider the situation that these CP maps have the same properties.
The environment is assumed to be in the vacuum state since thermal photons can be neglected in optical frequency region.
Under these assumptions, the CP maps $\hat{\cal L}^A$ and $\hat{\cal L}^B$ are given by \cite{walls1994}
\begin{equation}
\label{eq-2-3}
	\mathcal{L}^{A}=\exp\left[g\left(\hat{\mathcal{K}}_{-}^{A}
	-\hat{\mathcal{K}}_{0}^{A}+\frac{1}{2}\right)\right]
	\quad\quad\quad
	\mathcal{L}^{B}=\exp\left[g\left(\hat{\mathcal{K}}_{-}^{B}
	-\hat{\mathcal{K}}_{0}^{B}+\frac{1}{2}\right)\right],
\end{equation}
where $g$ is a positive parameter and the superoperators $\hat{\cal K}^A_-$ and $\hat{\cal K}^A_0$ are defined by
\begin{equation}
\label{eq-2-4}
	\hat{\mathcal{K}}_{-}^{A}\hat{X}
	=\hat{a}\hat{X}\hat{a}^{\dagger} \quad\quad\quad
	\hat{\mathcal{K}}_{0}^{A}\hat{X}
	=\frac{1}{2}(\hat{a}^{\dagger}\hat{a}\hat{X}
	+\hat{X}\hat{a}^{\dagger}\hat{a}+\hat{X}),
\end{equation}
for an arbitrary operator $\hat{X}$, and $\hat{\cal K}^B_-$ and $\hat{\cal K}^B_0$ follow the same definitions.
The CP maps $\hat{\cal L}^A$ and $\hat{\cal L}^B$ transform a coherent state into another coherent state with a reduced complex amplitude such as
\begin{equation}
\label{eq-2-5}
	\hat{\mathcal{L}}\vert\alpha\rangle\langle\beta\vert
	=E(\alpha,\beta)\vert\alpha\sqrt{T}\rangle\langle
	\beta\sqrt{T}\vert,
\end{equation}
where $T= \exp (-g)$ and $E(\alpha,\beta)$ is the function
\begin{equation}
\label{eq-2-6}
	E(\alpha,\beta)=\exp\left[-\frac{1}{2}(1-T)\left(
	\vert\alpha\vert^{2}+\vert\beta\vert^{2}-2\alpha\beta^{*}
	\right)\right].
\end{equation}
The parameter $T$ represents the transmittance of the noisy quantum channel.
Although this is one of the simplest loss mechanism in quantum channels, it can model experimental situations well.
A variety of other loss schemes was discussed in \cite{scheel2000}.

Suppose that Alice has an arbitrary quantum state $\hat{\rho}^C_{\rm in}$ which is to be teleported to Bob's hand.
The operator $\hat{\rho}^{ABC}=\hat{\rho}^{AB}_{\rm SV} \otimes \hat{\rho}^C_{\rm in}$ represents the total quantum state of Alice and Bob.
To teleport the quantum state $\hat{\rho}^C_{\rm in}$, Alice performs the simultaneous measurement of the position and the momentum of the mode A and C \cite{leon1997} described by the projection operator $\hat{X}^{AC} (x,p) = |\Phi^{AC} (x,p) \rangle \langle \Phi^{AC} (x,p) |$.
The vector $|\Phi^{AC} (x,p) \rangle$ is the simultaneous eigenstate of $\hat{x}^C - \hat{x}^A$ and $\hat{p}^C + \hat{p}^A$,
\begin{eqnarray}
\label{eq-2-7}
	\fl \vert\Phi^{AC}(x,p)\rangle=\frac{1}{\sqrt{2\pi}}
	\int_{-\infty}^{\infty}\rmd y\,\vert x^{C}+y^{C}\rangle
	\otimes\vert y^{A}\rangle \exp(\rmi py) \nonumber\\ 
	\lo=\frac{1}{\sqrt{2\pi}}\exp\left[-\frac{1}{2}\vert\mu\vert^{2}
	-\frac{1}{2}\rmi px+\mu\hat{c}^{\dagger}-\mu^{*}\hat{a}^{\dagger}
	+a^{\dagger}c^{\dagger}\right]\vert 0^{A}\rangle
	\otimes\vert 0^{C}\rangle,
\end{eqnarray}
where $\hat{c}$ and $\hat{c}^{\dagger}$ are the bosonic annihilation and creation operators of the mode C and $\mu = (x+ip)/\sqrt{2}$.
The probability $P(x,p)$ that Alice obtains the measurement outcome $(x,p)$ is given by
\begin{equation}
\label{eq-2-8}
	P(x,p)=\Tr_{ABC}\left[\left(\hat{X}^{AC}(x,p)\otimes\hat{1}^{B}
	\right)\left(\hat{\rho}_{\mathrm{SV}}^{AB}\otimes\hat{\rho}^{C}_{\rm in}
	\right)\right].
\end{equation}
Alice informs Bob of her measurement outcome $(x,p)$ by a classical communication channel.
By using the state-reduction formula \cite{davies1976,kraus1983,ozawa1984}, the quantum state $\hat{\rho}^B (x,p)$ at Bob's hand becomes
\begin{equation}
\label{eq-2-9}
	\hat{\rho}^{B}(x,p)=\frac{\Tr_{AC}\left[\left(\hat{X}^{AC}(x,p)
	\otimes\hat{1}^{B}\right)\left(\hat{\rho}_{\mathrm{SV}}^{AB}
	\otimes\hat{\rho}^{C}_{\rm in}\right)\right]}
	{\Tr_{ABC}\left[\left(\hat{X}^{AC}(x,p)\otimes\hat{1}^{B}
	\right)\left(\hat{\rho}_{\mathrm{SV}}^{AB}\otimes\hat{\rho}^{C}_{\rm in}
	\right)\right]}.
\end{equation}
After receiving the Alice's measurement outcome $(x,p)$, Bob applies the unitary operator $\hat{D}^B (x,p)= \rme^{i(p\hat{x}^B-x\hat{p}^B)}= \rme^{\mu\hat{b}^{\dagger}-\mu^* \hat{b}}$ to the quantum state $\hat{\rho}^B (x,p)$ and he finally obtains 
\begin{equation}
\label{eq-2-10}
	\hat{\rho}^{B}_{\mathrm{out}}(x,p)
	=\hat{D}^{B}(x,p)\hat{\rho}^{B}(x,p)\hat{D}^{B\,\dagger}(x,p).
\end{equation}
Averaging the output $\hat{\rho}^B_{\rm out}(x,p)$ over the probability distribution of $P(x,p)$ in \Eref{eq-2-8}, the averaged output state for Bob $\hat{\rho}^B_{\rm out}$ is derived as
\begin{eqnarray}
\label{eq-2-10.1}
\hat{\rho}^B_{\rm out} = 
\int^{\infty}_{-\infty} dx \int^{\infty}_{-\infty} dp 
\, P(x,p) \hat{\rho}^B_{\rm out} (x,p)
\nonumber\\ 
\fl= \int^{\infty}_{-\infty} \rmd x \int^{\infty}_{-\infty} \rmd p \,
\hat{D}^{B}(x,p) \Tr_{AC}\left[\left(\hat{X}^{AC}(x,p)
	\otimes\hat{1}^{B}\right)\left(\hat{\rho}_{\mathrm{SV}}^{AB}
	\otimes\hat{\rho}^{C}_{\rm in}\right)\right] \hat{D}^{B\,\dagger}(x,p).
\end{eqnarray}

\section{Completely positive map representation}

\begin{figure}
\begin{center}
\includegraphics[scale=0.7]{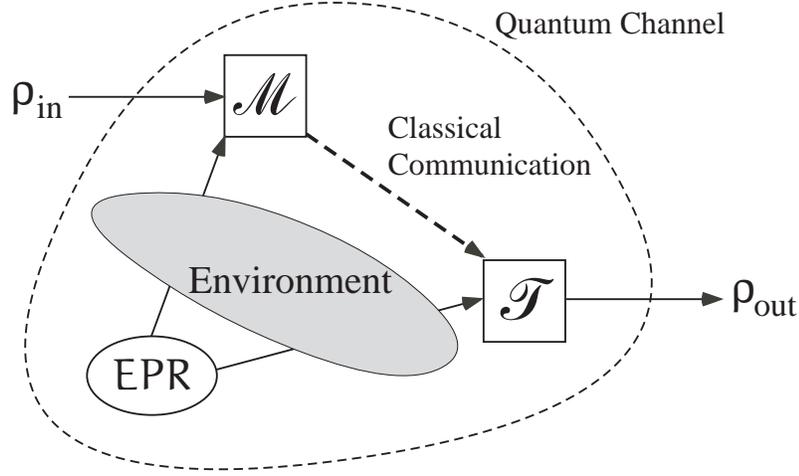}
\end{center}
\caption{Quantum channel of the teleportation.
In the figure, `M' stands for the quantum measurement performed
at the sender side, `T' represents the unitary transformation
carried out at the receiver side and `EPR' indicates
the entangled quantum state shared by Alice and Bob.}
\label{fig-channel}
\end{figure}

In this section, we derive the CP map representation for the continuous variable quantum teleportation described in the previous section.
A schematic of the channel is depicted in \Fref{fig-channel}.
The channel consists of the following procedure; the simultaneous measurement of the position and the momentum at the input side, the classical communication of the simultaneous measurement outcome, and the unitary transformation at the output side.
Since the simultaneous measurement is automatically performed inside the channel, the users (the sender and the receiver) of this channel do not care about its outcome while the simultaneous measurement gives different values of the position $x$ and the momentum $p$ with the probability $P(x,p)$ for each transmission as mentioned above.
Thus the output quantum state of this channel $\hat{\rho}^B_{\rm out}$ should be given as an average of each teleportation output $\hat{\rho}^B_{\rm out} (x,p)$, that is equal to the averaged output state discussed in the last section (\Eref{eq-2-10.1}).

Using Equations (\ref{eq-2-1}), (\ref{eq-2-5}) and (\ref{eq-2-7}) and the help of the completeness of the coherent states of each mode, the teleportation process of the input state $\hat{\rho}^C_{\rm in}$ is calculated as
\begin{eqnarray}
\label{eq-2-10.2}
\fl\Tr_{AC}\left[\left(\hat{X}^{AC}(x,p)
	\otimes\hat{1}^{B}\right)\left(\hat{\rho}_{\mathrm{SV}}^{AB}
	\otimes\hat{\rho}^{C}_{\rm in}\right)\right]
\nonumber\\
=\int \frac{\rmd^2 \alpha_1}{\pi} \int \frac{\rmd^2 \alpha_2}{\pi}
\int \frac{\rmd^2 \beta_1}{\pi} \int \frac{\rmd^2 \beta_2}{\pi}
\langle \alpha^A_1, \beta^B_1 | \Psi^{AB}_{\rm SV} \rangle
\langle \Psi^{AB}_{\rm SV} | \alpha^A_2, \beta^B_2 \rangle
\nonumber\\
\quad  \langle \Phi^{AC} | \Big( 
	{\cal L}^A |\alpha^A_1\rangle\langle\alpha^A_2| 
	\otimes \hat{\rho}^C_{\rm in} \Big) | \Phi^{AC} \rangle 
	\otimes {\cal L}^B | \beta^B_1 \rangle \langle \beta^B_2 |
\nonumber\\
= {\cal L}^T \Big[ D^{\dagger} (x,p) \hat{\rho}^C_{\rm in} D (x,p) \Big],
\end{eqnarray}
where $|\alpha, \beta \rangle = |\alpha\rangle \otimes |\beta\rangle$.
The CP map ${\cal L}^T$ is given by 
\begin{eqnarray}
\label{eq-2-10.3}
\fl{\cal L}^T \hat{X} = 
\int^{\infty}_{-\infty} \rmd y_1 \int^{\infty}_{-\infty} \rmd y_2
\int^{\infty}_{-\infty} \rmd z_1 \int^{\infty}_{-\infty} \rmd z_2 \,
G(y_1, y_2, z_1, z_2) 
| z_1 \rangle\langle y_1 | \hat{X} | y_2 \rangle\langle z_2 |,
\end{eqnarray}
where
\begin{eqnarray}
\label{eq-2-10.4}
\fl G(y_1, y_2, z_1, z_2)=
\frac{1}{2\pi^2 \sqrt{\bar{n}_{\lambda T} \bar{n}_{\lambda T}}}
\exp \left[ -\frac{1-\lambda^2 T^2}{
	(1-\lambda^2)^2 \bar{n}_{\lambda T} \bar{n}_{-\lambda T}} \bigg\{ \right.
\nonumber\\
\fl \qquad 
	  \frac{1}{4} (1+\lambda T)^2 F(z_1-y_1,z_2-y_2)
	+ \frac{1}{4} (1-\lambda T)^2 F(z_1+y_1,z_2+y_2)
\nonumber\\
\fl \qquad \left. \left.
	- \frac{1}{2} \lambda^2 (1-T)^2 F(y_1,y_2)
	- \frac{1}{2} \lambda^2 (1-T)^2 F(z_1,z_2)
	\right\} \right], \\
\fl F(x,y) = \left[ x + \frac{\lambda(1-T)}{1+\lambda T} y \right]
	\left[ x - \frac{\lambda(1-T)}{1-\lambda T} y \right]
	+ \left[ y + \frac{\lambda(1-T)}{1+\lambda T} x \right]
	\left[ y - \frac{\lambda(1-T)}{1-\lambda T} x \right],
\end{eqnarray}
and 
\begin{equation}
	\bar{n}_{\lambda T}=1-\frac{2\lambda T}{1+\lambda}
	=1-\left(1-\rme^{-2r}\right)T.
\label{eq-2-10.5}
\end{equation}
The quantum state $\hat{\rho}^B_{\rm out} (x,p)$, which is received at Bob's hand for each time teleportation, is then given by 
\begin{equation}
\label{eq-2-10.6}
\hat{\rho}_{\rm out}^B (x,p) =
\frac{D(x,p) {\cal L}^T \Big[ 
	D^{\dagger}(x,p) \hat{\rho}^C_{\rm in} D(x,p) \Big]
	D^{\dagger}(x,p)}{
	\Tr_B \{ {\cal L}^T [ D^{\dagger}(x,p) \hat{\rho}^C_{\rm in} D(x,p) ] \} },
\end{equation}
and finally the averaged output $\hat{\rho}_{\rm out}^B$ in \Eref{eq-2-10.1} is reformulated as
\begin{equation}
\label{eq-2-10.7}
\hat{\rho}_{\rm out}^B = 
\int^{\infty}_{-\infty} \rmd x \int^{\infty}_{-\infty} \rmd p \,
D(x,p) {\cal L}^T 
\Big[ D^{\dagger}(x,p) \hat{\rho}^C_{\rm in} D(x,p) \Big] 
D^{\dagger}(x,p).
\end{equation}
In the limit of the noiseless channel, the function $G(y_1, y_2, z_1, z_2)$ is divided as
\begin{equation}
\label{eq-2-10.8}
\lim_{T \to 1} G(y_1, y_2, z_1, z_2) = G(z_1, y_1) G(z_2, y_2),
\end{equation}
where
\begin{equation}
\label{eq-2-10.9}
G(x,y) = \frac{1}{\pi \sqrt{2}}
	\exp \left[ -\frac{1}{4} \rme^{2r} (x-y)^2 
	-\frac{1}{4} \rme^{2r} (x+y)^2 \right].
\end{equation}
In this case, \Eref{eq-2-10.3} together with Equations (\ref{eq-2-10.8}) and (\ref{eq-2-10.9}) corresponds to the position-momentum basis description of the CP map composed of the transfer operator derived in \cite{hofmann2000}.

\section{Formulation of the continuous variable quantum teleportation based on the $P$-function representation}

An arbitrary quantum state can be represented in a diagonal form with respect to coherent states, which is the $P$-function representation \cite{walls1994}.
In the following section, we reformulate the CP map of the teleportation channel in the last section into the $P$-function representation to derive the fidelity and the nonclassical depth.
The reformulation of the teleportation channel based on the $P$-function representation provides us some physical insights and the most straightforward formulation in order to quantify the transmission of the nonclassicality of the input quantum states with respect to the nonclassical depth.
It is well known that when the $P$-function is singular or not positive definite, the quantum state is nonclassical.
In addition, it is easier to obtain the density operator of the teleported quantum state from the $P$-function representation compared to other probabilistic function basises such as the Wigner function representation.

The $P$-function representation of the arbitrary input quantum state $\hat{\rho}^C_{\rm in}$ is given by 
\begin{equation}
\label{eq-2-11}
	\hat{\rho}^{C}_{\rm in}=\int \rmd^{2}\alpha\,
	P_{\mathrm{in}}(\alpha)\vert\alpha^{C}\rangle
	\langle\alpha^{C}\vert,
\end{equation}
where the $P$-function $P_{\rm in} (\alpha)$ is related to its density operator by the following formula:
\begin{equation}
\label{eq-2-12}
	P_{\mathrm{in}}(\alpha)=\frac{1}{\pi}
	\rme^{\vert\alpha\vert^{2}}\int \rmd^{2}\beta\,
	\langle-\beta^{C}\vert\hat{\rho}^{C}_{\rm in}\vert\beta^{C}\rangle\,
	\rme^{\vert\beta\vert^{2}}\rme^{\alpha\beta^{*}
	-\alpha^{*}\beta}.
\end{equation}
It is clear from \Eref{eq-2-11} that
if the teleported output state for a coherent state input is found,
the teleported quantum state for an arbitrary input is automatically given.
After tedious calculations, we obtain
\begin{eqnarray}
	\fl\hat{D}^{B}(x,p)\Tr_{AC}\left[\left(\hat{X}^{AC}(x,p)
	\otimes\hat{1}^{B}\right)\left(\hat{\rho}_{\mathrm{SV}}^{AB}
	\otimes\hat{\rho}^{C}_{\rm in}\right)\right]\hat{D}^{B\,\dagger}(x,p)
	\nonumber\\
	\lo=\frac{1-\lambda^{2}}{2\pi^{2}\lambda^{2}T(1-T)}
	\int\rmd^{2}\alpha\int\rmd^{2}\beta\,P_{\mathrm{in}}(\alpha)
	\vert\beta^{B}\rangle\langle\beta^{B}\vert
	\nonumber\\
	\lo\quad\quad\quad\quad\quad\quad\times
	\exp\left[-\frac{1-\lambda^{2}}{\lambda^{2}T}
	\vert\beta-\mu\vert^{2}-\frac{1}{1-T}\left\vert
	\frac{\beta-\mu}{\lambda}+\mu-\alpha\right\vert^{2}\right]
	\nonumber\\
	\lo=\frac{1-\lambda^{2}}{2\pi[1-\lambda^{2}(1-T)]}
	\int\rmd^{2}\alpha\,P_{\mathrm{in}}(\alpha)
	\hat{D}^{B}(\mu_{\alpha\lambda T})
	\hat{\rho}_{\tilde{n}_{\lambda T}}^{B}
	\hat{D}^{B\,\dagger}(\mu_{\alpha\lambda T}) \nonumber\\
	\lo\quad\quad\quad\quad\quad\quad\times
	\exp\left[-\frac{1-\lambda^{2}}{1-\lambda^{2}(1-T)}
	\vert\alpha-\mu\vert^{2}\right],
\label{eq-2-13}
\end{eqnarray}
where the parameters $\mu_{\alpha\lambda T}$ and $\tilde{n}_{\lambda T}$
are given by
\begin{equation}
	\mu_{\alpha\lambda T}=\mu+\frac{\lambda T}{1-\lambda^{2}(1-T)}
	(\alpha-\mu) \quad\quad\quad
	\tilde{n}_{\lambda T}=\frac{\lambda^{2}T(1-T)}{1-\lambda^{2}
	(1-T)}.
\end{equation}
In \Eref{eq-2-13}, the density operator
$\hat{\rho}_{\tilde{n}_{\lambda T}}^{B}$ represents
the thermal state with average photon number
$\tilde{n}_{\lambda T}$, that is,
\begin{equation}
	\hat{\rho}_{\tilde{n}_{\lambda T}}^{B}
	=\frac{1}{1+\tilde{n}_{\lambda T}}\sum_{n=0}^{\infty}
	\left(\frac{\tilde{n}_{\lambda T}}{1+\tilde{n}_{\lambda T}}
	\right)^{n}\vert n^{B}\rangle\langle n^{B}\vert.
\end{equation}
The probability $P(x,p)$ that Alice obtains
the measurement outcome $(x,p)$ is found to be
\begin{equation}\fl
	P(x,p)=\frac{1-\lambda^{2}}{2\pi[1-\lambda^{2}(1-T)]}
	\int \rmd^{2}\alpha\, P_{\mathrm{in}}(\alpha)
	\exp\left[-\frac{1-\lambda^{2}}{1-\lambda^{2}(1-T)}
	\vert\alpha-\mu\vert^{2}\right],
\label{eq-2-14}
\end{equation}
which clearly shows that the probability becomes
uniform in the strong squeezing limit $(\lambda\to1)$.
After receiving the measurement outcome $(x,p)$ from Alice, Bob performs the unitary transformation.
Then the teleported quantum state $\hat{\rho}_{\mathrm{out}}^{B}(x,p)$
reads
\begin{equation}
	\fl\hat{\rho}_{\mathrm{out}}^{B}(x,p)
	=\frac{\int\rmd^{2}\alpha\,P_{\mathrm{in}}(\alpha)
	\exp\left[-\frac{1-\lambda^{2}}{1-\lambda^{2}(1-T)}
	\vert\alpha-\mu\vert^{2}\right]
	\hat{D}^{B}(\mu_{\alpha\lambda T})
	\hat{\rho}_{\tilde{n}_{\lambda T}}^{B}
	\hat{D}^{B\,\dagger}(\mu_{\alpha\lambda T})}
	{\int \rmd^{2}\alpha\, P_{\mathrm{in}}(\alpha)
	\exp\left[-\frac{1-\lambda^{2}}{1-\lambda^{2}(1-T)}
	\vert\alpha-\mu\vert^{2}\right]}.
\label{eq-2-15}
\end{equation}
In the strong squeezing limit ($\lambda\to1$), the quantum state
$\hat{\rho}_{\mathrm{out}}^{B}(x,p)$ reduces into
\begin{equation}
	\hat{\rho}_{\mathrm{out}}^{B}(x,p)
	=\int\rmd^{2}\alpha\,P_{\mathrm{in}}(\alpha)
	\hat{D}^{B}(\alpha)\hat{\rho}_{1-T}^{B}
	\hat{D}^{B\,\dagger}(\alpha),
\label{eq-2-16}
\end{equation}
where the density operator $\hat{\rho}_{1-T}^{B}$ is the thermal state
with the average photon number $1-T$.
It is found from \Eref{eq-2-16} that in the strong squeezing limit,
the teleported quantum state $\hat{\rho}_{\mathrm{out}}^{B}(x,p)$
does not depend on the measurement
outcome $(x,p)$ obtained by Alice.
Furthermore, in the noiseless quantum channel $(T=1)$,
the teleported quantum state $\hat{\rho}_{\mathrm{out}}^{B}(x,p)$
becomes identical with the quantum state $\hat{\rho}^{C}_{\rm in}$
that Alice prepared.

As mentioned above, the final quantum state $\hat{\rho}_{\mathrm{out}}^{B}(x,p)$ at Bob's hand is distributed with the probability $P(x,p)$.
Thus the teleported quantum state averaged over this probability distribution is given by
\begin{eqnarray}
	\hat{\rho}_{\mathrm{out}}^{B}
	&=\int_{-\infty}^{\infty}\rmd x
	\int_{-\infty}^{\infty}\rmd p\,P(x,p)
	\hat{\rho}_{\mathrm{out}}^{B}(x,p) \nonumber\\
	&=2\int\rmd^{2}\mu\,P(x,p)
	\hat{\rho}_{\mathrm{out}}^{B}(x,p) \nonumber\\
	&=\frac{1}{\pi[1-(1-\rme^{-2r})T]}
	\int\rmd^{2}\alpha\int\rmd^{2}\beta\,P_{\mathrm{in}}(\alpha)
	\vert\beta^{B}\rangle\langle\beta^{B}\vert  \nonumber\\
	&\quad\quad\quad\quad\quad\quad\quad\quad\quad \times
	\exp\left[-\frac{\vert\alpha-\beta\vert^{2}}
	{1-(1-\rme^{-2r})T}\right]
	\nonumber\\
	&=\int\rmd^{2}\alpha\,P_{\mathrm{in}}(\alpha)
	\hat{D}^{B}(\alpha)\hat{\rho}_{\bar{n}_{\lambda T}}^{B}
	\hat{D}^{B\,\dagger}(\alpha),
\label{eq-2-17}
\end{eqnarray}
where the density operator $\hat{\rho}_{\bar{n}_{\lambda T}}^B$
represents the thermal state with average photon number
$\bar{n}_{\lambda T}$ which is given in \Eref{eq-2-10.5}.
For example, the output sate for the coherent state input $P_{\mathrm{in}}(\beta) = \delta^{(2)}(\alpha-\beta)$ is given by 
$	\hat{\rho}_{\mathrm{out}}^{B}
	=\hat{D}^{B}(\alpha)
	\hat{\rho}_{\bar{n}_{\lambda T}}^{B}
	\hat{D}^{B\,\dagger}(\alpha)$
which represents the thermal coherent state.
More generally, the basis of the coherent state expansion ($|\alpha \rangle \langle \alpha|$) in the input state $\hat{\rho}^C_{\rm in}$ is transformed into the thermal coherent state $\hat{D}(\alpha) \hat{\rho}_{\bar{n}_{\lambda T}} \hat{D}^{\dagger}(\alpha)$ by the lossy channel teleportation.
Here we note that the environment in the teleportation channel does not directly degrade the transferred quantum state $\hat{\rho}_{\rm in}$.
Degradation of $\hat{\rho}_{\rm in}$ in this teleportation model is caused by the imperfect squeezing and the environmental decoherence of the two-mode squeezed-vacuum before the displacement operation by Bob.
As a consequence, the losses of the teleportation channel injects the thermal noise into the transferred quantum state with the average photon number $\bar{n}_{\lambda T}$ instead of the direct degradation of $P_{\rm in}(\alpha)$ in \Eref{eq-2-17}.
It characterizes the teleportation channel definitely different from the direct transmission channel as discussed later.

The teleportation channel is also characterized by the following transformation of the $P$-function;
\begin{equation}
	P_{\mathrm{in}}(\alpha)\,\longrightarrow\,
	P_{\mathrm{out}}(\alpha)=\frac{1}{\pi\bar{n}_{\lambda T}}
	\int\rmd^{2}\beta\,P_{\mathrm{in}}(\beta)
	\exp\left(-\frac{\vert\alpha-\beta\vert^{2}}{\bar{n}_{\lambda T}}
	\right),
\label{eq-2-19}
\end{equation}
where $P_{\rm out}(\alpha)$ is the $P$-function of the quantum state $\hat{\rho}_{\rm out}$ and this transformation simply shows the thermalization process of the teleportation.

At the end of this section, we connect the $P$-function representation to the other quasi-probabilistic representations such as the Wigner function representation \cite{Braunstein1998,loock2000,Lee2000} by deriving a generalized quasi-probabilistic expression for the continuous variable teleportation.
The generalized $s$-ordered phase-space function $W^{(s)} (\alpha)$ \cite{Agarwal1970} is given by
\begin{eqnarray}
W^{(s)} (\alpha) &= \frac{1}{\pi(1-s)}
   \int \rmd^2 \beta \, P(\beta) 
   \exp \left( -\frac{|\alpha-\beta|^2}{1-s} \right)
\nonumber\\
   &=\frac{1}{\pi} {\rm Tr} \Big[ \hat{\Delta}^{(s)} (\alpha) \hat{\rho} \Big],
\label{eq-2-19-1}
\end{eqnarray}
where $0 \le s \le 1$ and the operator $\hat{\Delta}^{(s)} (\alpha)$ is defined as
\begin{equation}
\hat{\Delta}^{(s)} (\alpha) = \frac{1}{\pi}
   \int \rmd^2 \beta \, \hat{D} (\beta) 
   \exp \left[ \left( s-\frac{1}{2} \right) |\beta|^2
   - ( \alpha^{*} \beta - \alpha \beta^{*} ) \right].
\label{eq-2-19-2}
\end{equation}
It is clear that the $s$-ordered phase-space function $W^{(s)} (\alpha)$ is equal to the $P$-function, the Wigner function and the $Q$-function for $s=1$, $s=1/2$ and $s=0$, respectively.
It is found from Equations (\ref{eq-2-19}) and (\ref{eq-2-19-1}) that the continuous variable teleportation transforms the $s$-ordered phase-space function as follows:
\begin{equation}
W^{(s)}_{\rm in} (\alpha) \,\longrightarrow\, W^{(s)}_{\rm out} (\alpha)
=\frac{1}{\pi \bar{n}_{\lambda T}} \int \rmd^2 \beta \,
W^{(s)}_{\rm in} (\beta) \exp \left( 
-\frac{|\alpha-\beta|^2}{\bar{n}_{\lambda T}} \right),
\label{eq-2-19-3}
\end{equation}
which yields
\begin{equation}
W^{(s)}_{\rm out} (\alpha) = W^{(s-\bar{n}_{\lambda T})}_{\rm in} (\alpha).
\end{equation}
Thus the output state of teleportation is generally expressed as
\begin{eqnarray}
\hat{\rho}_{\rm out} &= \int \rmd^2 \alpha \, W^{(s)}_{\rm out} (\alpha)
    \hat{\Delta}^{(1-s)} (\alpha) \nonumber\\
    &=\int \rmd^2 \alpha \, W^{(s-\bar{n}_{\lambda T})}_{\rm in} (\alpha)
    \hat{\Delta}^{(1-s)} (\alpha).
\label{eq-2-19-4}
\end{eqnarray}

\section{Fidelity and the nonclassical depth}
\subsection{Fidelity}

For the teleportation of the pure input state $\hat{\rho}^{C}_{\rm in}=|\psi^{C}\rangle\langle\psi^{C}|$, the fidelity is given by 
\begin{eqnarray}
	F^{\rm tel}&=\langle \psi^{C} | \hat{\rho}^{B}_{\rm out} | \psi^{C} \rangle
	\nonumber\\
	&=\pi\int\rmd^{2}\alpha\,Q_{\mathrm{in}}(\alpha)
	P_{\mathrm{out}}(\alpha) \nonumber\\
	&=\frac{1}{\bar{n}_{\lambda T}}\int\rmd^{2}\alpha
	\int\rmd^{2}\beta\,Q_{\mathrm{in}}(\alpha)
	P_{\mathrm{in}}(\beta)
	\exp\left(-\frac{\vert\alpha-\beta\vert^{2}}
	{\bar{n}_{\lambda T}}\right),
\label{eq-2-23}
\end{eqnarray}
where $Q_{\mathrm{in}}(\alpha)=(1/\pi)\vert\langle
\alpha^{C}\vert\psi^{C}\rangle\vert^{2}$ is the $Q$-function
of the original state.
It is easy to see that
\begin{equation}
	\lim_{T\to1}\lim_{\lambda\to1}F^{\rm tel}=1 \quad\quad\quad
	\lim_{\lambda\to0}F^{\rm tel}=\pi\int\rmd^{2}\alpha\,
	Q_{\mathrm{in}}^{2}(\alpha)<1.
\label{eq-2-24}
\end{equation}
Naively the perfect quantum teleportation of continuous variables
is possible in the limit where the effect of the environment is negligible
and the squeezing is sufficiently strong.
When the original state is the coherent state,
the fidelity becomes
\begin{equation}
	F^{\rm tel}_{\rm coh} (\bar{n}_{\lambda T}) 
	 =\frac{1}{1+\bar{n}_{\lambda T}}
 	 =\frac{1+\lambda}{2(1+\lambda-\lambda T)}.
\label{eq-2-25}
\end{equation}
In case of the ideal quantum teleportation ($T=1$),
this result is identical with that obtained
so far \cite{hofmann2000}.
Similarly, the fidelities for the Fock state $|n\rangle$ and the Schr\"{o}dinger-cat state $(|\alpha \rangle - |-\alpha \rangle)/\sqrt{2(1-\rme^{-2|\alpha|^2})}$ are given by 
\begin{equation}
    F^{\rm tel}_{n} (\bar{n}_{\lambda T})
	=\frac{1}{1+\bar{n}_{\lambda T}}
	\left(\frac{1-\bar{n}_{\lambda T}}{1+\bar{n}_{\lambda T}}
	\right)^{n}P_{n}\left(\frac{1+\bar{n}_{\lambda T}^{2}}
	{1-\bar{n}_{\lambda T}^{2}}\right),
\label{eq-2-26}
\end{equation}
and
\begin{equation}
	F^{\rm tel}_{\rm cat} (\bar{n}_{\lambda T})
	=\frac{1}{2(1+\bar{n}_{\lambda T})}
	\left\{1+\left[\frac{\sinh\left(
	\frac{1-\bar{n}_{\lambda T}}{1+\bar{n}_{\lambda T}}
	\vert\alpha\vert^{2}\right)}
	{\sinh\left(\vert\alpha\vert^{2}\right)}
	\right]^{2}\right\},
\label{eq-2-27}
\end{equation}
respectively, where $P_{n}$ is the Legendre polynomial of order $n$.
It is worth noting that the fidelities in Equations (\ref{eq-2-25}), (\ref{eq-2-26}) and (\ref{eq-2-27}) take finite values even if $T=0$ (the input state is completely lost and turned into the vacuum state).
This is because the output state can still be made with the finite optical energy at Bob's hand after the classical communication from Alice.

\subsection{Nonclassical depth}

As mentioned above, the nonclassicality of quantum states
are very important in the fields of quantum optics 
and quantum information theory.
For example, it has recently been shown that the nonclassicality
is necessary to produce the entanglement of quantum states
by means of a beam splitter \cite{kim2001b}.
In this subsection, we briefly follow the definition of the nonclassical depth first, and then investigate the transfer property of the nonclassical depth by teleportation.
The nonclassical depth $\tau_{c}$ \cite{Lee1991,Lee1992,Lutkenhaus1995} of the quantum state
$\hat{\rho}$ is defined as
the minimum value of the parameter $\tau$ which
gives the non-negative value of the following quantity 
$R(\alpha,\tau)$ for all $\alpha$.
\begin{equation}
	R(\alpha,\tau)=\frac{1}{\pi\tau}\int\rmd^{2}\beta\,
	P(\beta)\exp\left(-\frac{\vert\alpha-\beta\vert^{2}}
	{\tau}\right),
\label{eq-3-1}
\end{equation}
where $P(\alpha)$ is the $P$-function of the quantum state
$\hat{\rho}$ and $\tau$ is a real parameter.
The mode superscripts are omitted until they are necessary.
It is clear that $R(\alpha,0)$, $R(\alpha,\frac{1}{2})$ and 
$R(\alpha,1)$ correspond to the $P$-function, the Wigner function
 and the $Q$-function of the quantum state, respectively.
Since the $Q$-function of an arbitrary state is non-negative,
the inequality $\tau_{c}\le1$ always hold.
The classical states provide $\tau_c = 0$ because its $P$-function
 is always non-negative without exception.
Thus the nonclassical depth $\tau_{c}$ satisfies the inequality
\begin{equation}
	0\le\tau_{c}\le1.
\label{eq-3-2}
\end{equation}
The Fock state and the superposition of two coherent
states (the Schr\"odinger-cat state) have the nonclassical depth
of unity ($\tau_{c}=1$).
The nonclassical depth of the single-mode squeezed state
with squeezing parameter $\xi=r \rme^{i\theta}$ is given by
\begin{equation}
	\tau_{c}=\frac{\tanh\vert\xi\vert}
	{1+\tanh\vert\xi\vert}=\frac{1}{2}\left(
	1-\rme^{-2\vert\xi\vert}\right),
\label{eq-3-3}
\end{equation}
which takes the maximum value of $\frac{1}{2}$
in the strong squeezing limit.
This result is understood from the fact that
the Wigner function of the single-mode squeezed state is non-negative.
Obviously, the coherent state and the thermal state
have the nonclassical depth of zero.

\begin{figure}
\begin{center}
\includegraphics[scale=0.8]{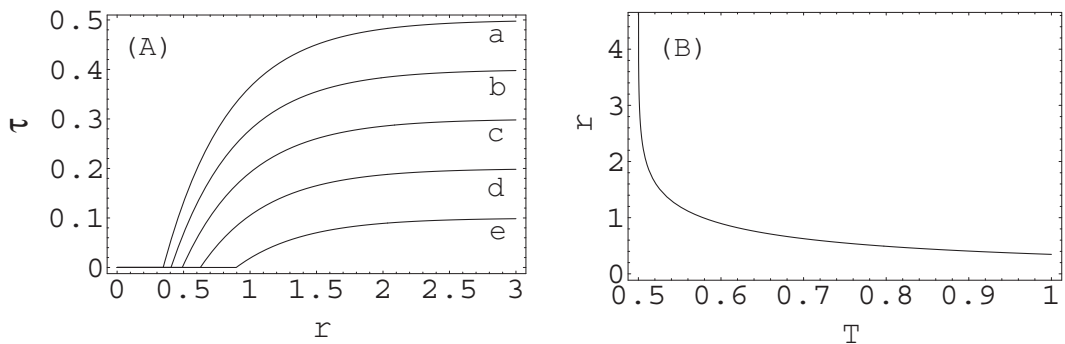}
\end{center}
\caption{Figure (A) shows the nonclassical depth $\tau_{\mathrm{out}}$
of the teleported quantum states $\hat{\rho}_{\mathrm{out}}$
in case of (a) $T=1.0$, (b) $T=0.9$, (c) $T=0.8$, (d) $T=0.7$
and (e) $T=0.6$. 
Figure (B) indicates the lower bound of the squeezing parameter
for obtaining the non-vanished nonclassical depth 
of the teleported quantum state $\hat{\rho}_{\mathrm{out}}$.
In the both figures, the nonclassical depth of the input quantum state
$\hat{\rho}_{\mathrm{in}}$ is assumed to be $\tau_{\mathrm{in}}=0.5$.}
\label{fig-depth}
\end{figure}

The nonclassical depth of the teleported quantum state
$\hat{\rho}_{\mathrm{out}}$ is easily found
from \Eref{eq-2-17}.
Substituting \Eref{eq-2-17} into \Eref{eq-3-1},
the $R$-function of the teleported quantum state
$\hat{\rho}_{\mathrm{out}}$ is calculated as
\begin{equation}
	R(\alpha,\tau)=\frac{1}{\pi(\tau+\bar{n}_{\lambda T})}
	\int\rmd^{2}\beta\,P_{\mathrm{in}}(\beta)
	\exp\left(-\frac{\vert\alpha-\beta\vert^{2}}
	{\tau+\bar{n}_{\lambda T}}\right).
\label{eq-3-4}
\end{equation}
This equation shows that the nonclassical depth 
of the original state $\tau_c^{\mathrm{in}}$ and that of 
the teleported state $\tau_c^{\mathrm{out}}$ are related
by the following relation:
\begin{equation}
	\tau_{c}^{\mathrm{out}}=\max\left[\tau_{c}^{\mathrm{in}}
	-\bar{n}_{\lambda T},0\right]
	=\max\left[\tau_{c}^{\mathrm{in}}-1
	+(1-\rme^{-2r})T,0\right].
\label{eq-3-5}
\end{equation}
Thus for the teleported quantum state $\hat{\rho}_{\mathrm{out}}$
to keep the nonclassical properties,
the squeezing parameter $r$ of the two-mode
squeezed-vacuum state shared by Alice and Bob have to satisfy
\begin{equation}
	r>-\frac{1}{2}\ln\left(1-\frac{1-\tau_{c}^{\mathrm{in}}}{T}
	\right).
\label{eq-3-6}
\end{equation}
If $T\le 1-\tau_{c}^{\mathrm{in}}$,
none of any nonclassical properties remain
in the teleported quantum state.
If $r=0$ (no squeezing), the nonclassical properties will never be teleported, {\it i.e.}, $\tau^{\rm out}_{c}=0$ as expected.
\Fref{fig-depth} shows the nonclassical depth $\tau^{\rm out}_{c}$ of the teleported state $\hat{\rho}^{\rm out}$ as a function of the squeezing parameter $r$ for several loss parameters $T$ (A), and the lower bound of $r$ for obtaining finite $\tau^{\rm out}_{c}$ at Bob's hand (B).
\Eref{eq-3-5} implies that
the transmitting performance measured by the nonclassical depth explicitly does not depend on what kind of nonclassical state is to be teleported.
It means that if two quantum states have the same nonclassical depth,
their teleported quantum states also have
the same nonclassical depth, 
independently of their quantum properties.
For example, the Fock state and the Scr\"odinger-cat state have
the same nonclassical depth $(\tau_{c}=1)$ and thus have 
the same transfer properties about the nonclassicality while
the fidelities in \Eref{eq-2-26} and \Eref{eq-2-27} are obviously different.

Before comparing the results of this section to that of the direct transmission channel in the next section, we mention about two kinds of possibilities to transmit quantum states by teleportation.
When the distance between Alice and Bob is $n$ times longer than the previous model, that is $\rme^{-g}=T \to \rme^{-ng}=T^n$, which is better to adopt one teleportation by the channel with the transmittance $T^n$ or the $n$ times iterative teleportations with the transmittance $T$ for each teleportation channel?
The answer is derived from \Eref{eq-2-19}.
For the one time teleportation with the transmittance $T^n$, the average photon number for thermalization is straightforwardly obtained as $\bar{n}_{\lambda T}(T^n) = 1-(1-\rme^{-2})T^n$ where $\bar{n}_{\lambda T}(T^n) \le 1$ for all natural number $n$.
For the latter case, $P_{\rm out} (\alpha)$ for the $n$ times iterative teleportations is given by the $n$ times convolutions of \Eref{eq-2-19} as
\begin{equation}
P_{\rm out}^{(n)} (\alpha) = \frac{1}{\pi} \frac{1}{n \bar{n}_{\lambda T}}
\int \rmd^2 \beta P_{\rm in} (\beta) \exp \left( 
   -\frac{|\alpha-\beta|^2}{n \bar{n}_{\lambda T}} \right),
\label{eq-3-7}
\end{equation}
where the effective average photon number for thermalization is given by $\bar{n}_{\lambda T}^{(n)} = n(1-(1-\rme^{-2})T)$ which linearly depends on $n$.
We obviously find that $\bar{n}_{\lambda T}(T^n) \le \bar{n}_{\lambda T}^{(n)}$ for all $r$, $T$ and natural number $n$ and thus recognize that a sequential connection of more than two teleportation channels is not necessary for a one terminal communication.

\section{Teleportation and the direct transmission}

In this section, we discuss the fidelity and the nonclassical depth of the continuous variable teleportation channel and that of the direct transmission channel.
It is reasonable to assume that the direct transmission channel is defined by the CP map which is used to transfer the two-mode squeezed-vacuum in the teleportation channel (\Eref{eq-2-3}).
By this assumption, the CP map of the direct transmission ${\cal L}$ is characterized by the transmittance $T$.
In the above sections, it has been tacitly assumed that the source of the two-mode squeezed-vacuum is located on the middle point of the whole teleportation channel and the two quantum channels for the two-mode squeezed-vacuum have the same length.
Thus the CP maps ${\cal L}^{A,B} (T)$ and ${\cal L} (T^2)$ should be used as the channels for the two-mode squeezed-vacuum and the channel for the direct transmission, respectively.
In the following section, we first derive the formula of the fidelity and the nonclassical depth in the direct transmission channel for given $T$ and then compare them with those of the teleportation channel with appropriate lengths.

With the help of \Eref{eq-2-5}, the output state for the direct transmission channel with the transmittance $T$ is calculated as
\begin{equation}
\label{eq-4-1}
\hat{\rho}^{\rm dir}_{\rm out} = \frac{1}{T} \int d^2 \alpha \,
P_{\rm in} \left( \frac{\alpha}{\sqrt{T}} \right) 
| \alpha \rangle \langle \alpha |,
\end{equation}
where the input state is given by \Eref{eq-2-11}.
The transmission fidelity is also derived in a same manner of the above sections as
\begin{equation}
\label{eq-4-2}
F^{\rm dir} = \frac{\pi}{T} \int d^2 \alpha \, 
Q_{\rm in} (\alpha) P_{\rm in} \left( \frac{\alpha}{\sqrt{T}} \right).
\end{equation}
Equations (\ref{eq-4-1}) and (\ref{eq-4-2}) clearly show that the $P$-function of the input state after the direct transmission is directly degraded by the environment while the loss of the teleportaion is the thermalization process as shown in Equations (\ref{eq-2-17}) and (\ref{eq-2-23}).
Actually the fidelities for the Fock state and the Schr\"{o}dinger-cat state transmissions by the direct transmission calculated from \Eref{eq-4-2} are written by the function of $T$ as 
\begin{equation}
\label{eq-4-2.1}
F^{\rm dir}_{\rm cat} (T) = \left[ 
\frac{\sinh (\sqrt{T}|\alpha|^2)}{\sinh (|\alpha|^2)} \right]^2
\cosh \big( (1-T)|\alpha|^2 \big),
\end{equation}
and
\begin{equation}
\label{eq-4-2.2}
F^{\rm dir}_{n} (T) = \exp [ n \log T ] = e^{-gn},
\end{equation}
respectively.
These are obviously different from Equations (\ref{eq-2-26}) and (\ref{eq-2-27}).
\begin{figure}
\begin{center}
\includegraphics[scale=0.7]{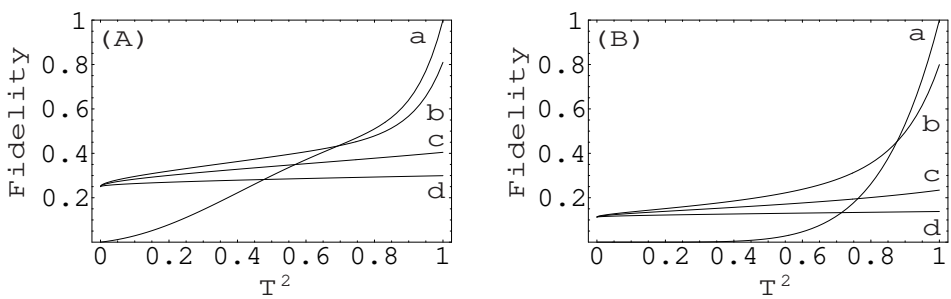}
\end{center}
\caption{Dependence on the transmittance $T^2$ of (A) $F_{cat}$ and (B) $F_{n}$  for (a) the direct transmission and the teleportation with (b) $r=2.0$, (c) $r=0.7$ and (d) $r=0.2$, respectively, where $\bar{n}=|\alpha|^2=6.0$.}
\label{fig-fidelity}
\end{figure}
Dependence on $T$ of the fidelities for the teleportation $F^{\rm tel}_{\rm cat} (\bar{n}_{\lambda T}(r,T))$ and $F^{\rm tel}_{n} (\bar{n}_{\lambda T}(r,T))$ for several $r$ are compared to those of the direct transmission $F^{\rm dir}_{\rm cat} (T^2)$ and $F^{\rm dir}_{n} (T^2)$ in \Fref{fig-fidelity}.

The $R$ function for the nonclassical depth of the directly transmitted state is also given by
\begin{equation}
\label{eq-4-3}
R^{\rm dir} (\alpha,\tau) = \frac{1}{\pi\tau} \int d^2 \beta \,
P_{\rm in} (\beta) 
\exp \left( -\frac{|\sqrt{T}\beta-\alpha|^2}{\tau} \right),
\end{equation}
where $T$ is the transmittance of the direct transmission channel.
By considering the definition of the nonclassical depth and $R^{\rm dir} (\alpha \sqrt{T}, \tau)$, the nonclassical depth for the output state of the direct transmission $\tau^{\rm dir}_{\rm out}$ is simply derived as
\begin{equation}
\label{eq-4-4}
\tau^{\rm dir}_{\rm out} (T) = \tau_{\rm in} T.
\end{equation}
The channel always transmit a part of $\tau_{\rm in}$ when the input state is nonclassical.

\begin{figure}
\begin{center}
\includegraphics[scale=0.7]{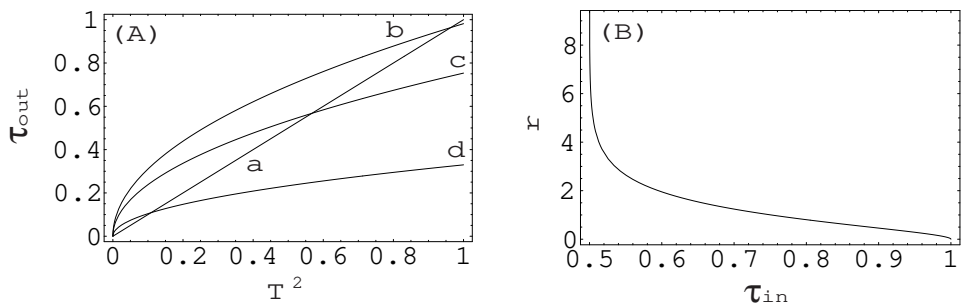}
\end{center}
\caption{Figure (A) shows the dependence of the nonclassical depth $\tau_{out}$ on the transmittance $T$ for (a) the direct transmission and the teleportation with (b) $r=2.0$, (c) $r=0.7$ and (d) $r=0.2$, respectively, where $\tau_{\rm in}=1.0$. 
Figure (B) indicates the lower bound for existing the positive $\tau_{D}(T)$ within $0 \le T \le 1$.}
\label{fig-noncl}
\end{figure}

Since the nonclassical depth of the outputs for the teleportation and direct transmission channels are simply given by Equations (\ref{eq-3-5}) and (\ref{eq-4-4}), respectively, it is now able to compare them analytically.
Define the difference between the two kinds of quantum channels $\tau_{\rm D} (T)$ as
\begin{eqnarray}
\label{eq-4-5}
\tau_{\rm D} (T) & = & 
\tau^{\rm tel}_{\rm out} (r,T) - \tau^{\rm dir}_{\rm out} (T^2)
\nonumber \\ & = &
- \tau_{\rm in} T^2 + (1-\rme^{-2r}) T + \tau_{\rm in} - 1
\qquad (0 \le T \le 1).
\end{eqnarray}
The bound for the positive $\tau_{\rm D} (T)$ is easily found as
\begin{equation}
\label{eq-4-6}
(1-\rme^{-2r})^2 > 4\tau_{\rm in} (1-\tau_{\rm in}).
\end{equation}
When Inequality (\ref{eq-4-6}) is fulfilled, the region for the positive $\tau_{\rm D} (T)$ is given by
\begin{eqnarray}
\label{eq-4-7}
&& \frac{(1-\rme^{-2r}) - 
\sqrt{(1-\rme^{-2r})^2+4\tau_{\rm in}(\tau_{\rm in}-1)}}{2\tau_{\rm in}}
\nonumber\\ &&
\qquad < T < \frac{(1-\rme^{-2r}) + 
\sqrt{(1-\rme^{-2r})^2+4\tau_{\rm in}(\tau_{\rm in}-1)}}{2\tau_{\rm in}}.
\end{eqnarray}
Since $T$ given in Inequality (\ref{eq-4-7}) must fulfill $0 \le T \le 1$ simultaneously, $\tau_{\rm in}$ in Inequalities (\ref{eq-4-6}) and (\ref{eq-4-7}) has the condition of
\begin{equation}
\label{eq-4-7.5}
\frac{1}{2} \le \tau_{\rm in} \le 1.
\end{equation}
Dependence on $T$ for nonclassical depths of $\tau^{\rm tel}_{\rm out} (r,T)$ and $\tau^{\rm dir}_{\rm out} (T^2)$ and the bound (\ref{eq-4-6}) for $1/2 \le \tau_{\rm in} \le 1$ are illustrated in \Fref{fig-noncl}.
For the input state with the maximal nonclassicality $\tau_{\rm in} \to 1$, Inequality (\ref{eq-4-7}) is simplified to
\begin{equation}
\label{eq-4-8}
0 < T < (1-\rme^{-2r}),
\end{equation}
and it gives the positive $\tau_{\rm D} (T)$ for all $T$ in the strong squeezing limit ($r \to \infty$).

These results show that the better choice to transmit the nonclassicality of the quantum states still depends on the loss of the quantum channel and the nonclassical depth of the transferred quantum state itself.
For more generalized characterizations, further quantum information theoretical approaches will be necessary.

\section{Concluding remarks}

In conclusion, we have formulated the quantum teleportation of continuous variables in noisy channels by using the CP map and the $P$-function.
It is assumed that the noise is due to the coupling between the system and the vacuum field.
The transmission performance is investigated with respect to the fidelity and the nonclassical depth.
These are compared to those of the direct transmission channel and we find that the teleportation is better than the direct transmission in a certain region.
Physically decoherence mechanisms are different between these two channels.
In teleportation channel, the loss is caused by an imperfect entanglement of the two-mode squeezing and the interaction between an environment and the two-mode squeezed-vacuum state instead of the quantum state $\hat{\rho}_{\rm in}$ itself.
So the decoherence in this case is effectively described by a thermalization process with the averaged photon number $\bar{n}_{\lambda T}=1-(1-\rme^{-2r})T$ although the loss model is assumed to be the interaction with a vacuum environment.
Mathematically, the losses of the fidelity and the nonclassicality in teleportation channel are given by convolutions of the $P$-functions for the input state $\hat{\rho}_{\rm in}$ and the thermal state $\hat{\rho}_{\bar{n}_{\lambda T}}$  while the losses in the direct transmission channel are described by degradations of amplitudes in the coherent state expansion.

We finally apply our results to the realistic situations in experiments.
In the experiment of \cite{furusawa1998}, the fidelity of $F=0.58 \pm 0.02$ was achieved for the coherent state teleportation with the amplitude efficiency of 0.9 for each two-mode squeezed-vacuum delivery.
Although the maximum squeezing of 6~dB was already observed at the frequency of $\Omega/2\pi=1.4$~MHz in the same experimental setup \cite{Polzik1992}, the teleported states generated by sideband modulations were prepared at the frequency of $\Omega/2\pi=2.9$~MHz where the effective squeezing was only 3~dB \cite{furusawa1998,loock2000}.
This is because there was technical noise at low modulation frequency.
Substituting $T=0.81$ and the effective squeezing of 3~dB ($r=0.34$) into \Eref{eq-2-25}, we obtain $F=0.62$ for the coherent state teleportation.
Although the fidelity obtained is slightly overestimated, our theoretical result shows a reasonable agreement.

Now we consider a transmission of the 6~dB squeezed state ($r=0.69$) by the teleportation channel with the 6~dB two-mode squeezed-vacuum.
The nonclassical depth of the 6~dB squeezed state corresponds to $\tau^{\rm in}_{c}=0.38$.
Thus, from \Eref{eq-3-5}, we find that the transmittance of $T>0.83$ is at least necessary for obtaining the nonzero nonclassical depth $\tau^{\rm out}_{c}$ after the teleportation.
It is also shown in \Fref{fig-noncl}~(B) that such teleportation channel will be better than the direct transmission channel with $T^2$ for the purpose of transferring highly nonclassical states, such as Fock states.

As shown in this paper, the performance of the continuous variable teleportation as a quantum communication channel depends not only on the parameters of the channel, but also on what kind of quantum information we want to send.
To find the best way for transmitting quantum information through various possible channels is generally a nontrivial problem.
It would be an important future problem to study efficient codings against some practical noise models in both teleportation and direct transmission scenarios.

\end{document}